\documentclass[sigconf, authorversion=true, nonacm=true]{acmart}
\AtBeginDocument{%
  \providecommand\BibTeX{{%
    \normalfont B\kern-0.5em{\scshape i\kern-0.25em b}\kern-0.8em\TeX}}}

\usepackage{xac}

\begin{document}

\title{GANzilla: User-Driven Direction Discovery in Generative Adversarial Networks}


\author{\textbf{Noyan Evirgen}}
\affiliation{%
  \institution{UCLA HCI Research}
  \country{}}
\email{nevirgen@ucla.edu}

\author{\textbf{Xiang `Anthony' Chen}}
\affiliation{%
  \institution{UCLA HCI Research}
  \country{}}
\email{xac@ucla.edu}

\newcommand{\gz}[0]
{{\sc GANzilla}\xspace}

\begin{abstract}
Generative Adversarial Network (GAN) 
is  widely adopted in numerous application areas, such as data preprocessing, image editing, and creativity support.
However, GAN's `black box' nature prevents non-expert users from controlling what data a model generates, spawning a plethora of 
prior work that focused on {\it algorithm-driven} approaches to extract editing directions to control GAN.
Complementarily, we propose a \gz---a {\it user-driven} tool that empowers a user with the classic scatter/gather technique to iteratively discover directions to meet their editing goals.
In a study with 12 participants, \gz users were able to discover directions that \one edited images to match provided examples (closed-ended tasks) and that \two met a high-level goal, \eg making the face happier, while showing diversity across individuals (open-ended tasks).

\blfootnote{To appear in UIST 2022}
\end{abstract}



\begin{teaserfigure}
\centering
  \includegraphics[width=0.85 \textwidth]{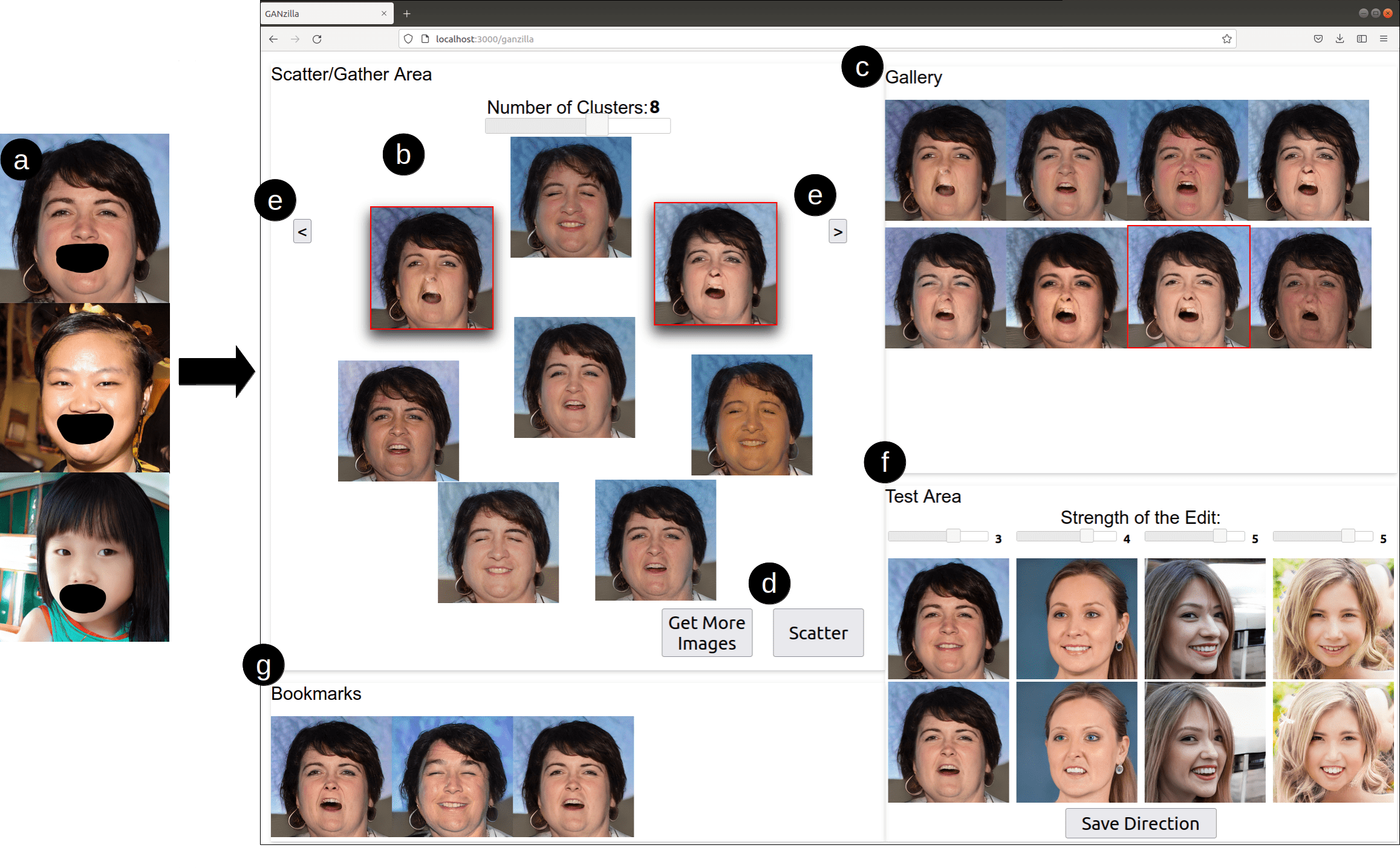}
  \caption{\gz is a tool that allows users to discover editing directions in Generative Adversarial Networks (GAN) via iterative scatter/gather interactions---a user-driven approach that complements many existing algorithm-driven methods. (a) A user starts by highlighting a region of interest (an optional step). (b) Based on the highlight (if there is), directions are sampled and clustered, each shown as an image edited by that direction. The user can gather clusters by selecting thumbnail images (indicated by a red border). (c) The user can see all the directions of the gathered clusters and (d) scatter them into new clusters. (e) The user can go back-and-forth across iterations to explore alternate choices of scatter/gather. (f) The user can test a selected direction (red border in c) on other images with individual sliders controlling the strength to apply the direction. (g) The user can bookmark directions that meet their editing goals.
  }
  \label{fg:fig1}
\end{teaserfigure}

\maketitle

\section{Introduction}
Generative Adversarial Networks (GANs) promise to create new content by learning the characteristics of existing data, showing compelling results in various domains, from stylization \cite{zhu2017unpaired}, scene creation \cite{vondrick2016generating}, and improving the quality of scientific data \cite{sandfort2019data}.

Unfortunately, despite its increasing widespread use, most GAN models to date remain a `black box' to end-users with little transparency about what a model is capable of generating and little control over the generative process. Without transparency and control, when end-users have a creative intent (e.g., a caricaturist illustrating a character’s facial expression in creative storytelling), they cannot see whether or how one can instruct a GAN model as-is to generate specific characteristics.
In other words, there is little support for formulating user-defined {\it editing directions}\footnote{Hereafter simply referred to as `direction'.}---a set of input parameters that steer the GAN model to generate contents with varying levels of characteristics (\eg making the hair on the input image more or less blonde).

To address the limited transparency and control, some prior work allows a user to browse GAN-generated results in an interactive gallery view \cite{10.1145/3411764.3445714}; however, such an open-ended exploration is not intended to converge to a specific direction.
Others propose methods to dissect a GAN model \cite{bau2018gan} or to perform post hoc extraction of principal components \cite{harkonen2020ganspace} or semantic controls \cite{Collins_2020_CVPR}. 
However, such directions are often pre-defined by algorithms, which do not permit a user to specify directions to generate their own desired characteristics.

We design and implement \gz---a tool that complements existing algorithm-driven approaches by enabling user-driven direction discovery in a GAN model.
As a proof of concept, we focus on a common use case of stylizing faces based on the StyleGAN2 model \cite{Karras2019stylegan2}; however, the workflow in our tool is expected to generalize to other usages of GAN as well.

As shown in \fgref{fig1}a, a \gz user starts with brushing on a few exemplar images demonstrating specific areas they want to stylize, based on which the back-end samples a large number of directions.
Then, \gz's front-end employs the classic scatter/gather technique \cite{pirolli1996scatter} to let a user iteratively filter directions\footnote{In \gz, a direction is represented as a thumbnail of an exemplar edited image, \eg an `aging' direction is shown as a face older than the original image.} applied on exemplar images and iteratively narrow down to the ones that result in their desired characteristics.
Specifically, a user 
can gather one or more clusters (\fgref{fig1}b), see whether the constituent directions interest them (\fgref{fig1}c), and, if so, scatter (\fgref{fig1}d) them into new clusters to refine their selection.
Flexibly, at any given time, the user can go back to the previous iterations and scatter a different subset of the clusters (\fgref{fig1}e). 
Meanwhile, the user can test directions on some other images and see if the effects generalize (\fgref{fig1}f) and bookmark the ones they like (\fgref{fig1}g).

We validate \gz in a user study ($N=12$) with two types of tasks:
\one In the closed-ended tasks, we controlled what a user intended to edit by providing participants a set of edited image pairs (references and targets). 
Results show that participants were able to find the GAN directions that closely replicated the edits---specifically, their discovered directions transformed the reference images into ones that are more similar to the target images and such similarities rank high when compared to edits done by 1000 randomly-sampled directions (representing the latent space).
\two In the open-ended tasks, we tested whether participants could use \gz to achieve personalized edits: given some high-level editing goals (\eg making the face happier), each participant would use \gz to find specific edits that they considered to achieve such goals.
Results show that each participant felt satisfied with their edits, which also highly align with the goals when analyzed from a language perspective (comparing image and text embeddings); further, the resultant edits exhibit diversity across participants, indicating \gz's ability to enable personalized content creation.


Overall, \gz makes a {\bf tool contribution}: in contrast to using a GAN model as a `black box', we provide a comprehensive tool for a user to see what directions the GAN model is capable of and to iteratively discover and test directions that generate their desired characteristics.
\gz's user-driven direction discovery aims to complement (rather than replace) existing algorithm-driven approaches \cite{Goetschalckx_2019_ICCV, Shen_2020_CVPR, yang2021semantic} by providing users with an option to explore more editing options when the pre-computed controls do not fully meet their needs. 

\section{Background \& Related Work}
In this section, we first provide background information on GAN and editing directions, then we review two areas of work that intersect with ours: existing algorithm-driven approaches for discovering GAN directions and enabling users to interact with GAN.





{\bf How does GAN work?}
As shown in \fgref{gan}, GAN is a family of neural networks where a generator $G$ is trained to create synthetic data ($x_g$) that simulates those from a certain domain (\eg images of human faces), a discriminator $D$ is trained to distinguish between the generator's synthetic data ($x_g$) and real data ($x_r$), and the countering of $D$ and $G$ iteratively leads to the generator's ability to create synthetic data indistinguishable from real data.
By default, GAN functions as a `black box' where a large collection of data points (\eg images) are generated from randomly-sampled `noises' ($z$), leaving users very little control of the generative process.

{\bf What is a GAN (editing) direction?}
The input `noise' ($z$ in \fgref{gan}) to the generator is a point from GAN's latent space, which typically consists of a high-dimension of variables, each drawn from a Gaussian distribution \cite{brownlee_2020}.
An editing direction is a vector $d$ in the latent space along which we can move $z$ so that the resultant $x_g = G(z)$ will change in a semantically meaningful way, \eg, given a closed-mouth image $G(z_0)$, making the mouth open in the image $G(z_1)$ where $z_1 = z_0 + \lambda d$. The coefficient $\lambda$ is the strength to apply direction $d$.


{\bf Algorithm-driven approaches for discovering GAN directions}.
To discover GAN directions, early work employs supervised approaches that require sampling and labeling a large number of points in the latent space \cite{jahanian2019steerability, Goetschalckx_2019_ICCV, Shen_2020_CVPR, plumerault2020controlling, yang2021semantic}. 
Recently, there has been a plethora of research focused on unsupervised methods \cite{pmlr-v119-voynov20a, Shen_2021_CVPR}, each of which aims at controlling a specific aspect of the generated data.
Härkönen \etal describe a PCA-based approach to decompose a GAN model into interpretable controls for users to specify desired attributes of the generated outcome \cite{harkonen2020ganspace}.
%
Wu \etal present a method for computing style channels, each controlling a distinct, localized visual attribute without entangling with one another
\cite{Wu_2021_CVPR}.
Collins \etal enable local semantic editing using GAN: by selecting a specific part of an image (\eg Person A's nose), GAN is able to transfer its style to another image (\eg making Person B's nose look like A's)
\cite{Collins_2020_CVPR}.
%
Although all the above work does enable high-level user-control of the GAN's output, such fully algorithm-driven approaches tend to result in one-size-fits-all directions and do not permit user input to specify customized directions.
Complementarily, \gz enables each individual user to discover directions on their own, which provides a useful option when the pre-computed controls do not fully meet a user's needs.

\fg{gan}{gan}{1}{Overview of a typical GAN model.}

{\bf Enabling users to interact with GANs}.
Even prior to the popularity of GAN, researchers have explored interactions with other generative processes, most of which are centered around generative design (\eg using topology optimization) via sketching \cite{10.1145/3126594.3126662, chen2018forte} or visualization \cite{10.1145/3173574.3173943} to explore a large design space.
%
Given that most conventional ways of controlling GAN is via the use of sliders, Dang \etal conducted a comprehensive comparative study to understand the effects of regular sliders \vs sliders that provide a `filmstrip' of feedforward information (preview images)
\cite{dang2022ganslider}.
Other researchers go beyond sliders to consider alternative techniques to interact with GAN.
For example, Zhang and Banovic present samples of generated images on a grid-like view wherein a user can zoom in/out or pivot to explore more images
\cite{10.1145/3411764.3445714}.
Alternatively, it is also possible to consider other input modalities beyond conventional GUI elements.
For example, Yu \etal demonstrate a visual design assistant that takes in a user's natural language feedback to guide the GAN model's modification of the design \cite{10.1145/3394171.3413551}.
Ling \etal allow users to discover directions by changing segmentation masks of the generated images \cite{ling2021editgan}. 
Related to our goal of enabling users to discover directions, Or \etal develop StyleCLIP that takes the input of textual description (\eg ``Mohawk hairstyle'') and transforms the image accordingly. However, this approach cannot accept arbitrary editing requests; rather, acceptable textual descriptions are limited to a set of phrases the CLIP model produces to characterize a given image dataset. In contrast, \gz does not limit a user to a pre-defined vocabulary but allows them to discover their desired editing direction by navigating vast examples using the scatter/gather technique.



\section{GANzilla: User-Driven GAN Direction Discovery for Image Editing}

In this section, we present a detailed walkthrough of \gz's design and implementation using an exemplar use case of stylizing a human portrait image to make the face happier.

\subsection{Highlighting an area to focus the edit on}
To start, the user can choose to select a specific part of the face for the edit to focus on. 
For example, the user might wish to edit the mouth to make a big smile so the face would appear happier.
To specify the mouth area, the user simply uses a built-in brush tool to paint over a few exemplar images \gz provides (\fgref{fig1}a).
Note that this step is optional: without any selection, edits could occur indiscriminatively across the entire image.

{\bf Implementation}. 
\gz runs the direction search on the StyleSpace of StyleGAN2. StyleSpace refers to the space that is defined by style parameters of StyleGAN2. These parameters control the individual strength of various filters in StyleGAN2. It is significantly more disentangled than the traditional latent space \cite{wu2021stylespace}. 
Depending on whether the user performs this optional highlighting step, the backend either uses the entire or a subset of the style parameters based on the highlighted region. We use a similar method to Collins \etal \cite{Collins_2020_CVPR} to translate a highlighted region into a subset of style parameters. Each style parameter of StyleGAN2 is assigned an importance metric depending on the highlighted region and the activation maps of the highlighted image. This step allows us to \textit{select} the filters of StyleGAN2. After doing this for few exemplar images, we take the union of the selected style parameters. These parameters are later used to sample directions.

\subsection{Sampling and clustering directions}
Next, based on the user's selected image region (if there is any), \gz samples a large number of directions, each of which is represented by an exemplar image edited along that direction.
\gz clusters the sampled directions and displays each cluster using one of its directions' thumbnail images (\fgref{sampling_clustering}b) to avoid crowding the UI.
Selecting one or multiple clusters shows the constituent directions in a separate view (\fgref{sampling_clustering}c).
The user can change the number of clusters (the default number is six) and if they cannot find a direction that matches their editing intent, they can request \gz to resample more directions (\fgref{sampling_clustering}d).

{\bf Implementation}. 
Sampling is done using the selected set of style parameters. We observed from StyleCLIP that the directions only needed a small number of style parameters to be diverse and expressive. Following that observation for each direction sample, we sub-sample the style parameters that come from the highlighting step. The sub-sampling rate is tuned as a hyper-parameter. The resulting set of style parameters are then randomly increased or decreased using a normal distribution, which defines the sampled directions. Each direction then changes a different set of style parameters to increase diversity. In order to cluster these directions, we first generate the resulting images. Then we extract latent codes for these images using an AI-model (CLIP model in our case). Next, these latent codes are clustered with the k-means clustering. Note that we can not use the directions directly without generating the images, because the dimensions of the directions are not comparable with each other. As for the representative image of the cluster, we choose the one that is closest to the center of the cluster. If the user asks for more directions, we repeat the sampling step while giving a higher priority to the style parameters that are not chosen in previously sampled directions. This allows us to investigate a new subspace in StyleSpace that is not covered in previous directions.

\fg{sampling_clustering}{sampling_clustering.png}{}{After highlighting, \gz generates an initial set of directions. (a) By default it returns six clusters of directions which are all displayed with one representative image each. (b) The user can change the number of clusters. (c) Selecting multiple clusters (indicated by a red border) shows the constituent images of those clusters. (d) If the user cannot find a direction, they can request to sample more.}

\fg{iterative}{iterative.png}{}{The resulting view after clusters gathered in \fgref{sampling_clustering} are scattered. (a) If the user feels unsatisfied with the scatter results, they can click the back button to go to the previous clusters. (b) The user can select a direction (indicated by red border) and (c) test how it works on other images: the first row are the reference images and the second row edited by the direction being tested, whose strength can be adjusted for individual test images. (d) The user can bookmark a direction if they are satisfied with its edits.  
}

\subsection{Iterative scatter/gather of directions}
Once a user identifies clusters of interest (\eg open-mouth images matching their intended editing goal), they can iteratively use the scatter/gather technique: first selecting those clusters (gather) and then clicking the `scatter' button to re-cluster them; then the user can repeat this process with more scatter/gather.
\fgref{iterative} shows the updated UI after scattering the two clusters gathered in \fgref{sampling_clustering}.
Here, \gz repurposes the classic scatter/gather technique \cite{pirolli1996scatter}, which was used for browsing a large collection of text documents, for enabling a user to iteratively converge their choices of direction.
Further, \gz allows a user to step back (to previous iterations) and explore alternate `branches' of scatter/gather.
Specifically, if the user feels unsatisfied with the current batch of directions, they can click the `<' button (\fgref{iterative}a), return to the previous clusters, gather and scatter a different subset of them.

{\bf Implementation}.
Scatter is implemented by first sampling new directions based on the gathered clusters and then clustering the new set of directions.
In order to sample a new direction, a random pair is selected from all the gathered directions. Then these two directions are averaged, creating a new direction that combines the two. In order to increase the variation, a random vector (sampled from a normal distribution) is added to the resulting direction. Results of scatters are stored in a tree data structure. When the user wants to explore alternate branches, a new subtree is created from the current node. 

\subsection{Testing directions on more images}

At any given time, the user can test a direction by selecting its thumbnail (\fgref{iterative}b)
and see how this direction works on other images. Each of these test images comes with a slider for the user to adjust the strength of the direction (how strongly to apply its editing effects) (\fgref{iterative}c). Such a `test field' allows a user to calibrate how strongly they should apply a direction and examine the direction's generalizability to make sure that it can generate the intended edits on other images as well. If a user is satisfied with the direction after the tests, they can bookmark it with the save button (\fgref{iterative}d). Later, the user can also bring back a 'bookmarked' direction to the test-area by clicking its thumbnail.


{\bf Implementation}. 
When a thumbnail image is clicked to be tested, \gz scales that direction with a default strength of the direction. Then, the direction is applied to different test images that can be uploaded by the user. The resulting images and their reference images are shown to the user. Whenever there is a change to a slider, the resulting image is re-generated.

\subsection{Other implementation details}
\label{other_implementation_details}
We used Pytorch as our deep learning backend. For the rest of the back-end implementation we used Python. We used Flask as our web-framework, which handled the communication between our front-end and the back-end. For our front-end, we used a combination of Javascript, Node.js and React. The back-end ran on a Linux server equipped with an Nvidia GeForce RTX 3090 GPU.








%








\section{User Study}
We conducted a study to validate whether \gz can enable users to discover directions that steer a GAN model to edit images for specific purposes.


\subsection{Participants}
We used convenience sampling to recruit $12$ participants from a local university (eight male, four female, aged 23 to 31).
Eight participants majored in electrical and computer engineering, one in bioengineering, one in medicine, one in mechanical engineering and one in engineering management.
Eleven participants had programming experiences ($3$ to $15$ years) and three had programmed or used GAN-enabled applications before (P5, P6 and P7)\footnote{We decided not to exclude these participants because \gz was meant for complementing existing algorithm-driven direction discovery and users who work on GAN development should be able to use and benefit from \gz as well.}.





\subsection{Tasks \& Procedure}

Each participant performed two blocks of editing task using \gz, each consisting of three trials.

\begin{itemize} [leftmargin=0.25in]
    \item {\bf Closed-ended tasks}. 
        In each trial, we provided a participant with a set of image pairs where each pair showed an image before and after editing (\fgref{closed_tasks}), which hereafter are referred to as {\it reference} and {\it target} images, respectively. The participant's goal was using \gz to find a direction that could replicate the edits, \ie transforming the reference images into ones that were as similar to the target images as possible.
    \item {\bf Open-ended tasks}.
        In each trial, we provided a participant with a set of images and a high-level editing goal---specifically, making all the faces old, happy, and surprised. The goals were intentionally open-ended so that the participant had to come up with specific edits based on their own interpretation of the goal and use \gz to discover directions accordingly. 
\end{itemize}


\fg{closed_tasks}{closed_tasks.png}{0.85}{Closed-ended tasks: the three sets of image pairs given to the participants where each column consists of a reference and a target images. A participant's task was to discover a direction that edits each reference image to best approximate the corresponding target image.  
}

Each study started with an introductory tutorial of \gz, followed by a brief practice session for each participant to try out \gz using a toy dataset.
We then continued with the block of open-ended tasks, 
after which the participant would take a short break before performing three trials of closed-ended tasks\footnote{In an earlier pilot study, participants found closed-ended tasks much more challenging; thus we always started with open-ended tasks to ease participants' learning curve.} (\fgref{closed_tasks}). 
The order of the three trials within each block were counter-balanced across participants.
Finally, we concluded the study with a semi-structured interview to elicit participants' qualitative feedback of interacting with \gz.
The entire study took place over Zoom and lasted for about one hour and each participant was compensated with a \$25 gift card.

\subsection{Data \& Apparatus}
For the back-end we used the state-of-the-art StyleGAN2 \cite{Karras2019stylegan2}. Together with its predecessor, StyleGAN \cite{karras2019style}, StyleGAN2 has been used for various applications including style transfer, data augmentation and image-editing.
For data we used the Flickr-Faces-HQ (FFHQ) dataset, which was originally created for the StyleGAN as a benchmark. As our deep learning model, we used a pretrained model that is trained on FFHQ and released by Nvidia for StyleGAN2 on their github page\footnote{https://github.com/NVlabs/stylegan2-ada-pytorch}. Other implementation details are introduced in \S~\ref{other_implementation_details}.
We conducted the study virtually where each participant used Zoom's remote desktop control to interact with the \gz front-end running in a Chrome Web browser on the experimenter's desktop computer.
The front- and back- ends were connected via a local area network to minimize latency.



\fgw{open_qual}{open_qual.png}{1}{Open-ended tasks. Participants were given a text description of the tasks (making the face old, happy, and surprised). The reference images are given at the first column. Rest of the columns are generated by directions discovered by participants. For the same editing goal, participants discovered a wide variety of directions using \gz. }

\fg{closed_ended_examples}{closed_ended_examples.png}{}{Closed-ended tasks. Participants were given the reference images (first column) and the target images (last column) to match as well as they can. The columns in the middle are generated by participants' discovered directions. 
}

\subsection{Measurement}
\label{section:measurement}
We recorded the entire Zoom meeting including the screen recording. In addition to that, we saved every image participant generated throughout the study. We also logged all of the user actions with timestamps including what they highlighted, which buttons they clicked, which directions they have tested and saved. 

For qualitative measures of open-ended tasks, immediately upon finishing each trial, we asked each participant to rate (along a 7-point Likert scale) how successful they thought they had achieved the editing goal with the direction they found.

In the exit interview, participants started with an overall assessment of \gz based on their overall experience for both closed- and open-ended tasks. We asked \one whether the tool is easy to use and \two whether the user can find directions that match their editing goal.
Next, participants rated the cognitive load using the mental demand, effort and frustration dimensions in the NASA TLX questionnaire \cite{hart1986nasa}.
Next, we asked participants to ablatively evaluate the usefulness of \gz's individual UI elements: highlighting, the scatter/gather technique, changing the number of clusters, going back-and-forth across iterations, asking for more images, and live-testing directions on multiple images. All questions were rated along a seven-point Likert scale.

\section{Quantitative Results}

\fgref{open_qual} and \fgref{closed_ended_examples} show sample images edited by participants' discovered directions for open- and closed-ended tasks, respectively.
Below we provide quantitative analyses to better understand participants' performance and behavior using \gz.




\subsection{Closed-ended tasks}
\label{closed_ended_quant_results}

\subsubsection{User performance}
We calculated the cosine similarity between the target image and a participant-generated image (\ie edited by the participant's discovered direction)
using VGG-Face's latent vector \cite{parkhi2015deep} extracted from the last layer. Cosine similarity results in between $0$ and $1$ where $1$ represents a closer match between the vectors.
Specifically, we ran two different analyses using this VGG-Face similarity metric.

\begin{table}[]
\begin{tabular}{lccc}
\toprule
               & Task 1 & Task 2 & Task 3 \\
\midrule
Reference      &    $0.388$     &    $0.252$    &    $0.377$    \\
User-Generated &    $0.446 \pm 0.123$    &     $0.422 \pm  0.112$   & $0.544 \pm 0.140$ \\
\bottomrule

\end{tabular}
\caption{
    Closed-ended tasks: user-generated images (edited by participants' discovered directions) are more similar to the target images across all three tasks compared to reference images. Higher value represents a closer match between the vectors.
}
\label{tab:closed_similarity}
\end{table}

In the first analysis, 
we tested whether a participant-generated image was {more similar} to the target image than the reference image is. 
As shown in Table~\ref{tab:closed_similarity}, 
the similarities between the participant-generated images and the target images (averaged across participants) were $0.446 \pm 0.123$, $0.422 \pm  0.112$ and $0.544 \pm 0.140$ for the three trials, respectively, 
all of which were higher than the similarities between the reference and the target images, which were $0.388$, $0.252$, and $0.377$. 
    
In the second analysis, we first sampled 1000 random directions to edit a reference image. We then calculated where a participant-generated image ranked amongst the 1000 randomly-edited images, in terms of their similarity to the target image. 
Results show that, in 33 out of 36 tasks (three tasks per participant $\times$ 12),
the participant-generated images ranked top-5,
which suggests that the participant-discovered directions got very close to the target image amidst the large GAN latent space (represented by the random samples).

\subsubsection{User behavior}
Overall, the average time to complete a closed-ended task is seven minutes and $16$ seconds. Participants spent $37.4\%$ of their time on performing scatter/gather interactions,
$45.0\%$ on testing a direction,
and $17.6\%$ on highlighting.

The average number of scatters per task is $1.64$ and for an average of $0.53$ times a participant went back to undo a scatter. 
We noted that some users scattered more than the others. Four participants contributed to almost half (49.2\%) of the total number of scatters.
The average number of directions tested per task is $6.61$. On average when these directions were being tested, the strength of the direction was changed $3.23$ times. 

On average, participants requested $1.81$ times per task to change the number of clusters, which ranged from six to nine across all tasks.
In comparison, participants only asked to sample more images $0.56$ times per task.

\subsection{Open-ended tasks}
\label{open_ended_quant_results}

\subsubsection{User performance}
We conducted three folds of analyses:

First, we employed an open-source face analyzing tool called Deepface \cite{serengil2021lightface}. We extracted the age and emotion predictions as well as their respective confidence levels.     
Using this tool, we can analyze whether a user-generated image (compared to the reference image) resulted in an increase in age (old) or the confidence level in emotions (happy and surprised). 
Results show that on average age increased by $10 \pm 3.12$ years and confidence values for happy and surprised increased by $45.1\% \pm 25.7\%$ and $56.7\% \pm 22.9\%$ respectively. 

    
Second, we used the CLIP model \cite{radford2021learning} that embeds both text and image into the same latent space to make them comparable. 
We first computed the {\it image-originated} embeddings, which correspond to each participant's discovered directions by performing a subtraction between the CLIP embedding of the participant-generated image and that of the reference image. 
Next, we computed the {\it text-originated} embeddings: for each editing goal, we used the corresponding keyword (`happy', `old', and `surprised') to create the text embeddings in an approach similar to StyleCLIP \cite{Patashnik_2021_ICCV}. 
%
%
We then compare the image- and text-originated embeddings by calculating their cosine-similarities, which were $0.314 \pm 0.131$, $0.341 \pm 0.192$ and $0.500 \pm 0.072$, respectively, for the three editing goals.

    


    
Third, to put these similarity numbers in perspective, we sampled 1000 random directions to represent the latent space and used the aforementioned Deepface tool to search for the top-10 directions that generated images with the highest increase in age/confidence in emotions.
We then compared the embeddings of these top-10 directions with the aforementioned text-originated embeddings.
We found that their averaged cosine-similarity values are $0.194 \pm 0.099$, $0.285 \pm 0.176$ and $0.387 \pm 0.172$, respectively, which are all lower than those achieved by user-generated images, as shown in Table~\ref{tab:open_similarity}.

\begin{table}[]
\begin{tabular}{lccc}
\toprule
               & Task 1 & Task 2 & Task 3 \\
\midrule
Top-10*     &    $0.194 \pm 0.099$    &    $0.285 \pm 0.176$    &    $0.387 \pm 0.172$     \\
User-Generated &    $0.314 \pm 0.131$  &  $0.341 \pm 0.192$     & $0.500 \pm 0.072 $\\
\bottomrule
\end{tabular}
*{\footnotesize Amongst 1000 random samples that approximate GAN's latent space} 
\caption{Open-ended tasks: Similarity of user-generated images and top-10 randomly generated images compared to the CLIP \textit{text-originated} embeddings. User-generated images are more similar to the given text (`old', `happy', and `surprised') across all tasks. Higher value represents a closer match between the vectors.}
\label{tab:open_similarity}
\end{table}

The three analyses above suggest that participants' discovered directions reached a high semantic proximity to the editing goal.
In addition, participants also felt positively about how they succeeded in achieving the editing goals, 
with reported scores (on a seven-point Likert scale) of
$6.08 \pm 0.76$, $5.83 \pm 0.99$ and $5.75 \pm 0.83$, respectively for the three tasks. 

\subsubsection{User Behavior}
We report the same set of behavioral measures as in the closed-ended tasks and compare the two, reporting statistical significance whenever there is any (otherwise any difference should be assumed as not reaching statistical significance).

Overall, the average time for a participant to complete an open-ended task is eight minutes and $55$ seconds, with $27.5\%$ of their time spent on scatter/gather,
$58.9\%$ on testing directions, and $13.6\%$ on highlighting. 
Participants used the test-field about two minutes more per task than the closed-ended tasks ($W=32, p=0.03$ based on a Wilcoxon signed-rank test), probably to ensure that the editing goal actually applied to more than one face for the open-ended tasks. 

The average number of scatters per task is $1.17$, about $0.5$ smaller than that in the closed-ended tasks ($W=8, p=0.02$).
Scattering more allows one to further refine a direction to better match the target image, which probably explained the higher number in the closed-ended tasks.
Unsurprisingly, the same participants who scattered more in closed-ended tasks also scattered more here, contributing to over half (54.8\%) of the total number of scatters.
On average, users went back $0.5$ times to undo scatter---a similar number as in the closed-ended tasks. 

The average number of directions tested per task is $7.11$.
When these directions were being tested, the strength of the direction was changed for an average of $3.67$ times. 
On average, participants requested $2.86$ times to change the number of clusters, which ranged from eight to ten. Participants only asked for more images $0.388$ times per task.

\begin{table}[]
    \resizebox{\columnwidth}{!}{%
    \includegraphics[width=1\columnwidth]{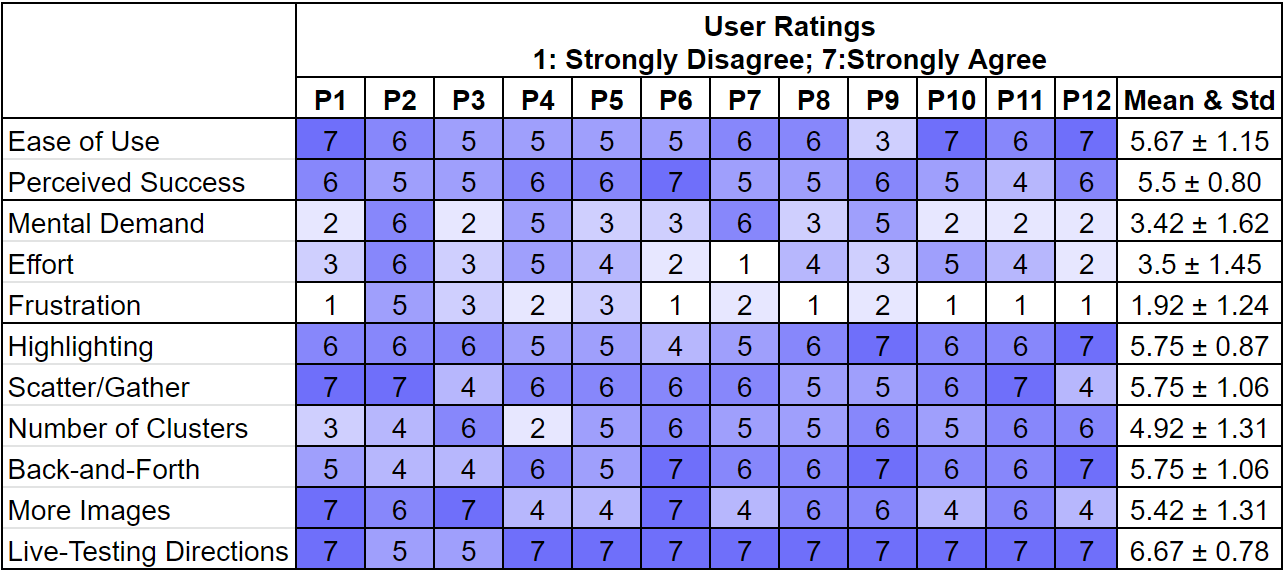}
    }
    \caption{The participants' ratings (Row 1-2: overall experience; 3-5: workload; the rest: usefulness of individual components). The questions are explained in \S~\ref{section:measurement}. All questions used a seven-point Likert scale. The outliers are found using interquartile range (IQR) analysis. P9 ease of use, P2 frustration, P4 number of clusters, P2 and P3 live-testing directions are outliers.}
    \label{tab:user_ratings}
    \end{table}




\subsection{Workload measured by NASA TLX}
For the mental demand dimension, the four participants (P2, P4, P7, and P9) who rated higher than four (neutral) considered the main workload as inspecting the small changes and differences amongst images during the scatter/gather process. 
As mentioned by P2, ``You need to take a look at the images, small changes in them. Then you have to envision what to combine which increases the mental load''. 
P2 is also the only participant who gave a higher-than-neutral rating of effort and the only one that rated frustration higher than three, which in part due to the need to spend a lot of time on some tasks because there were many options. Rating of P2 on frustration is an outlier based on the IQR analysis.
Similar concerns are shared by P1 and P4: ``Workflow is easy but still I had to click different images and pay attention. If there were no images I liked, it got harder because I had to start thinking about which images have the right parts to scatter'' (P1).
``Paying attention to the details can be demanding, focusing on the patterns as well'' (P4). P7 commented on the challenges of visualization of data: ``The task is really useful but visualizing really high dimensional data is challenging which increases the mental load'' (P7).

\subsection{Summary of quantitative results}

Analyses of the closed-ended tasks show that participants' discovered directions transformed the reference images into ones that are more similar to the target images and such similarities rank high when compared to edits done by 1000 randomly-sampled directions.

Analyses of the open-ended tasks show that participants' discovered directions achieve the given editing goals as validated by Deepface's age and emotion detection; further, such directions highly align with the interpretation of these goals from a language perspective.

Analyses of user behavior across both tasks show that 
\one a few (four) participants contributed to about half the total number of scatter/gather interactions;
\two participants scattered/gathered significantly more in closed-ended tasks; and
\three participants spent significantly more time testing directions in open-ended tasks.


Table~\ref{tab:user_ratings} shows participants' ratings of \gz with respect to ease of use, perceived success, workload, and an ablative assessments of each component's usefulness.
Next, we report participants' qualitative feedback behind these ratings.


\section{Qualitative Findings}
\label{qual}

We employed a method akin to the Affinity Diagram approach \cite{holtzblatt1997contextual}, based on which we aggregated participants’ responses to summarize their perceived ease and success of using our tool (\S \ref{overall}) and surfaced recurring themes regarding how participants assess the usefulness of \gz’s individual components (\S \ref{ablative}). Specifically, the first author transcribed participants’ responses to develop the initial codes, which were then reviewed by the second author. Disagreements were resolved via discussion between the two authors.

\subsection{Overall assessment}
\label{overall}

\subsubsection{Ease of using the tool}
When asked how \gz was easy to use, all but one (P9) participant gave a rating above five.
For example, P1 said: ``I don't have to remember most of the workflow. It is just highlight and then click on images based on what you are searching.'' 
Both P3 and P5 commented on the intuitiveness of the UI.
Some participants pointed out that there was a learning curve mainly due to the inevitable randomness of the sampling process (P3 and P6), \ie participants needed to learn how to develop a strategy of using \gz based on the sampled directions given to them.
P9 gave the only below-five rating and thought that it was hard to know how to improve the directions without getting overwhelmed by the sheer amount of information (``too many faces''). Rating of P9 is an outlier based on the IQR analysis.

\subsubsection{Perceived success of the tasks}
When asked how they felt successful that they found the directions to achieve their goals, all but one (P11) participant gave a rating above five.
Even P9 who did not feel \gz was easy to use considered the task successful: ``At the end of all tasks, I found a direction. They were not exact but close''.
Participants' responses also pointed out nuances between types of tasks---``Open-ended tasks were easy to achieve. Closed-ended were harder'' (P4) and
nuances between different stages of a task---``It is easy to find the main direction (bigger smile) but getting all the secondary changes are challenging'' (P7).
P11, who gave the only below-five rating, reflected on their usage strategy---``Sometimes I felt like it did not match the target very well. Maybe, I needed to iterate/scatter more. This is especially the case for closed-ended tasks'' (P11).

\subsection{Ablative assessment of individual components}
\label{ablative}
\subsubsection{Highlighting an area to focus the editing on presented limited usefulness} 
For open-ended tasks, sometimes participants did not know which part to edit before starting explorations, as mentioned by P1: ``It seems if I know exactly what I am looking for it is helpful. But there are some cases where it is not obvious so I can't imagine where to highlight.''
Perhaps a more noticeable issue of this component was GAN's entanglement problem (discussed in more details in \S~\ref{section:entanglement}), as participants noticed that sometimes the highlighted part was not guaranteed to be majorly edited (P2 and P4) and sometimes non-highlighted parts were also changed (P6 and P7).
Interestingly, one participant (P2) reported using this component, not to instruct the GAN model, but to remind themselves to focus on specific parts they wanted to edit.





\subsubsection{Scatter/gather helps to combine directions with different features}
Multiple participants (P1, P4, P5, P10, and P11) mentioned this usage, \eg ``There were a lot of cases that it was useful. For example I wanted to change both eyes and mouth but some clusters had only eyes and some clusters only had the mouth. I could leverage scatter to get both.'' (P1)
%
%
Participants also pointed the need for finer-grained gather, \eg 
``It would be nice if we could choose individual images instead of clusters. Because sometimes I did not want to choose the entire cluster.'' (P2)
Interestingly, one participant pictured scatter as ``zooming to that region'' (P12), which could inform them if the gathered directions ``are bad''.

\subsubsection{Changing the strength both positively and negatively in the testing field helps one better understand the direction}
The ability to live-test a direction on multiple images was rated the highest amongst all components.
Foremost, participants valued such a test of a direction's generality (P4, P5, and P6), as pointed out by P6: ``I could also see the directions on other images. So I have an idea about how well it works generally.''
Participants also realized the importance of exploring the right strength of applying a direction, \eg 
``The previous steps help me to find the direction but you still need to figure out the strength'' (P1),
``You can experiment with the magnitude of the vector. It allowed me to try different combinations and helped me build my intuition about the direction'' (P12).
To our surprise, many participants (P1, P4, P5, P6, P8, P10, and P12) heavily used the functionality of setting a negative value on the strength of the direction, which essentially allowed them to observe what happens if they go in the reverse direction. 
Specifically, seeing how a direction works in reverse was ``informative'' (P10), helped participants ``validate'' (P5) or ``understand'' (P4, P8, and P11) a direction better and ``convince'' (P12) themselves that it was the right direction.
P6 even employed a strategy that leveraged such negative strength: ``... in the last task I could not find the asked direction. Instead I found an opposite direction and used a negative weight on the test area.''
Amongst the two outlying scores (5), while P3 did not state anything specifically negative, P2 pointed out that he did not find testing highly useful because he could already anticipate how the direction would likely fail in some cases.

\subsubsection{Changing the number of clusters and requesting more samples help mitigate randomness}
Multiple participants (P1, P2, P5) pointed out the inherent randomness of sampling directions and considered that asking for more images was a back-up solution (P3, P6, P8, and P9) that helped when they could not find a direction that they were looking for. 
Sampling randomness also affected the quality of the clusters, as pointed out by a few participants (P1, P4, and P7).
Changing the number of clusters helped them make sense of the clusters.
For example, P12: ``It helped me to choose better groups because when you increase number of clusters, clusters become better refined''.
However, as pointed out by P1, one trade-off was having to track changes in the clusters as the number was changed.
P4 did not find changing the number of clusters useful because he ‘would rather see as many images’ as he could and thus always set the max number (10). Rating of P4 is an outlier based on the IQR analysis.
\section{Discussions \& Future Work}

We discuss several issues in the current system and possible solutions for future work.

{\bf Limitations of the current study}.
First, future work could increase the number of participants and the number of tasks (\eg via a out-of-lab deployment) beyond the current controlled study.
Second, three of our participants had prior knowledge of GAN. Although anecdotally we did not observe any difference in how these three participants used \gz, future work should still strive to focus on a narrower user group (\eg product designers who use GAN to formulate ideas).
Finally, to ease participants' learning curve, we fixed the order of the tasks to be open-ended (easier) first then closed-ended (harder). To verify whether there is an ordering effect, future work could extend our study with a counter-balanced design. For stylizing faces, mouth area is usually the most expressive. However, there were other changes in our closed-ended tasks. For example, Task 3 had more ‘squinty’ eyes and Task 2 had more makeup after the direction was applied. We do recognize such changes are more subtle compared to mouth and will address this limitation in our future work by introducing tasks that involve more significant changes in non-mouth areas.

{\bf Addressing entanglement issues in GANs}.
\label{section:entanglement}
Entanglement refers to a long-standing phenomenon in GAN direction discovery: if a feature is changed and another unintended feature is also changed, these two features are said to be \textit{entangled}. 
For example, while trying to make someone look happier, the image might also appear younger. In this example, the feature happy and young are entangled. In general, we want to discover directions that are disentangled. 

To mitigate this issue, we used the state-of-the-art StyleGAN2 and found the directions in StyleSpace which is significantly more disentangled than the latent space. 
However, entanglement still existed in our study and was pointed out by two participants: ``... when I try to make someone happy, their skin tone also changes'' (P4); ``I focus on the mouth but I get variety of eyes.'' (P5)
Interestingly, some participants used the scatter/gather technique to mitigate entanglement issues. 
For example, if the goal is to find a direction that results in a bigger mouth, by scattering a cluster that has the bigger mouth feature, all the entangled features in the cluster are scattered too. 
This allowed participants to choose a more disentangled direction. 

In the future, we can also preprocess the StyleSpace itself to address entanglement. Instead of randomly sampling various dimensions, we can be more selective about the sampling procedure. This can be achieved by uncovering dependencies between dimensions, so that related dimensions are selected together rather than entangled ones. Another idea is to let users inform the system about the entanglement issue in a secondary highlighting step. Then we can try to remove the StyleSpace parameters that are related with what user highlighted. An iterative process between the tool and the user can result in more disentangled directions, which is left for future work.

{\bf Providing users with more guidance and explanation}.
While \gz makes the generative process controllable as a whole, participants nonetheless requested more guidance and explanation on the specific steps in this process.
For example, participants wished the tool could provide some guidance when they were stuck (\ie faced with clusters that contained no directions related to their editing goal). 
One possible idea for future work, in this case, is to guide the user with a shortcut that jumps to a previous step that contains directions most different from the current ones.
Another popular suggestion is to help users to keep track of how clusters change and differ. For example, displaying a heatmap next to each image so that the changes/differences become more salient to human eyes.
One participant (P7) also suggested feedforward visualizations to help users preview what they will get if they perform certain actions.
Further, future work can introduce a recommender component that retrieves directions in previously-unsampled space based on what clusters a user currently gathers.

{\bf Integrating \gz with algorithm driven direction discovery}.
Currently, we do not leverage algorithm-driven directions and instead let the users discover them from scratch. Although this approach worked well in our studies, using prior work to discover directions can also be beneficial to the user experience. Some of the sampled directions can be filtered or better understanding of the editing space can be achieved. In the future, users can look through algorithm-driven directions and then can decide to use \gz if their needs are not satisfied. These directions can be analyzed by our back-end to improve \gz sampling and they can even become part of scatter/gather functionality. We consider GANzilla complementary to existing approaches \cite{harkonen2020ganspace, jahanian2019steerability} and thus did not compare their results. To address this, our future studies will incorporate state-of-the-art methods to show how user-driven directions might result in different edits than algorithm-driven approaches.\textbf{}


{\bf Improving scatter/gather with more engagement}
We observe that not all participants fully leverage scatter to `dive deep' into the StyleSpace but rather only scattered a few times. 
This is partly because it is challenging to manage highly-branched out scatter/gather paths as the number of iterations increases. 
Future designs could incorporate a tree-like UI structure of directions: 
every time the user scatters, a new branch is created. 
Another approach in the future can be to build the tree to guarantee hierarchical semantics. This can allow users to traverse a semantically more meaningful tree and and go `deeper' in to the StyleSpace for more specific directions. 
As it becomes easier to manage the scatter/gather iterations, we can further support exploring multiple directions at the same time (\eg making a mouth open and eye brows raised).

\begin{acks}
We thank the reviewers for their valuable feedback. 
We thank all the anonymous participants for their participation in our study. 
This work was funded in part by 
the National Science Foundation under grant IIS-2047297 and
the Office of Naval Research under grant N00014-22-1-2188.
\end{acks}

\bibliographystyle{ACM-Reference-Format}
\bibliography{ref_main}

\end{document}